\documentclass[graybox]{svmult}

\usepackage{mathptmx}       
\usepackage{helvet}         
\usepackage{courier}        
\usepackage{type1cm}        
\usepackage{amsfonts}                            
\usepackage{makeidx}         
\usepackage{graphicx}        
\usepackage{multicol}        
\usepackage[bottom]{footmisc}

\makeindex             

\newcommand{\nit}{\noindent}

\newcommand{\dsp}{\displaystyle}
\newcommand{\vs}[1]{\vspace{#1 ex}}
\newcommand{\hs}[1]{\hspace{#1 em}}
\newcommand{\bflr}{\begin{flushright}}
\newcommand{\eflr}{\end{flushright}}
\newcommand{\bc}{\begin{center}}
\newcommand{\ec}{\end{center}}
\newcommand{\ben}{\begin{enumerate}}
\newcommand{\een}{\end{enumerate}}

\newcommand{\be}{\begin{equation}}
\newcommand{\ee}{\end{equation}}
\newcommand{\ba}{\begin{array}}
\newcommand{\ea}{\end{array}}
\newcommand{\ct}{\cite}
\newcommand{\bit}{\bibitem}
\newcommand{\dd}[2]{\frac{\partial{#1}}{\partial{#2}}}

\newcommand{\del}{\delta}

\newcommand{\ve}{\varepsilon}

\newcommand{\thg}{\theta}
\newcommand{\kg}{\kappa}
\newcommand{\lb}{\lambda}
\newcommand{\sg}{\sigma}
\newcommand{\rg}{\rho}

\newcommand{\vf}{\varphi}
\newcommand{\og}{\omega}
\newcommand{\Gam}{\Gamma}
\newcommand{\Del}{\Delta}
\newcommand{\Fg}{\Phi}

\newcommand{\Og}{\Omega}

\newcommand{\bfk}{{\bf k}}

\newcommand{\bfr}{{\bf r}}

\newcommand{\bfx}{{\bf x}}

\newcommand{\uh}{\underline{h}}

\newcommand{\hr}{\hat{r}}

\newcommand{\lh}{\left(}
\newcommand{\rh}{\right)}
\newcommand{\ld}{\left.}

\newcommand{\nb}{\nabla}

\newcommand{\ddd}[1]{\stackrel{\cdots}{#1}}

\newcommand{\der}{\partial}

\begin{document}

\title*{Dynamical Space-Time and Gravitational Waves}
\author{J.W.\ van Holten}
\institute{J.W.\ van Holten \at Nikhef, Science Park 105, Amsterdam NL; \email{v.holten@nikhef.nl}}
%
%
\maketitle

\abstract{According to General Relativity gravity is the result of the interaction between matter and space-time geometry. 
In this interaction space-time geometry itself is dynamical: it can store and transport energy and momentum in the form 
of gravitational waves. We give an introductory account of this phenomenon and discuss how the observation of gravitational 
waves may open up a fundamentally new window on the universe.}

\section{Gravity \label{sec:1}}

One hundred years ago Albert Einstein published his first papers on a new theory of gravitational 
interactions: General Relativity (GR) \ct{einstein1915, einstein1916} . It is a marvelous achievement 
as it connected two hitherto completely separate braches of mathematics in one physical framework: 
field theory and geometry. It is such a rich theory that even hundred years later we have not yet 
worked out all aspects of it. Moreover in spite of many creative ideas and numerous efforts the 
question how it fits together with quantum theory has yet to find a completely convincing and 
satisfying answer. 

The connection between geometry and the physics of gravity can be approached from both sides. 
The physical approach is to start from elementary observations about free fall. Such observations 
suggest two principles underlying the nature of gravity: \\
a.\ {\em Universality:} all bodies are subject to gravity independent of their nature and composition. \\
b.\ {\em Equivalence principle:} all other circumstances being equal bodies of different mass fall with 
the same gravitational acceleration. \\
The equivalenc principle was established in the 16th century by an experiment with lead balls 
dropped from the tower of the New Church of Delft in the Netherlands by the flemish engineer and
mathematician Simon Stevin and his dutch friend Johan de Groot. Stevin published a description of
this experiment in his widely read book {\em Principles of the art of weighing} \ct{stevin1586}.

Both principles were embodied by Newton in his theory of gravity. Newton's law states that the 
gravitational force between two bodies depends only on their masses $m$ and $M$ and the relative 
distance $r = |\bfr_2 - \bfr_1|$, and is proportional to the product of the masses 
\[
F \propto \frac{m M}{r^2},
\]
whilst the acceleration of any one of the bodies, say that of mass $m$, is the ratio of the force and 
its mass
\[
a = \frac{F}{m}.
\]
Thus the acceleration depends on the mass $M$ of the other body which can be considered
the source of the gravitational force experienced by $m$, but not on that of the body falling 
toward $M$ itself. 

A weakness of Newton's theory, recognized by Newton himself, was that it required instantaneous 
action at a distance. Two centuries later it was replaced by Einstein's theory of General Relativity, which 
also incorporates the universality and equivalence principle but attributes them to the geometry of the 
space-time in which bodies are moving, not to some force acting instantaneously over arbitrary large
distances. This theory is build around the concept of a dynamical space-time, the properties of which
are determined by the masses and motion of the objects it contains. At the same time these motions
are constrained by the geometry of the space-time, making space-time itself the medium through which  
gravitational interactions between bodies take place.

\section{Field Theory \label{sec:2}}

Field theory was developed in the 19th century, mainly in connection with understanding the phenomena 
of electricity and magnetism. Based on the work by such people as Amp\`{e}re, Oersted and Faraday 
eventually Maxwell came up with a unified theory of electric and magnetic phenomena created by 
charges and currents, in which the idea of fields permeating all of space was essential. The appearance 
of a charge at some point in space created a local change in the field which propagated to other places
at large but finite speed, eventually influencing the motion of other charges. Maxwell's theory does not 
require instantaneous action at a distance, as all interactions between charges and fields are local.
Moreover Maxwell's theory has another feature which was only understood after great effort, mostly 
by the work of Lorentz: it incorporates special relativity as it is invariant under rotations and Lorentz 
transformations. Once Einstein recognized this to be a universal feature of physical phenomena it then 
motivated Minkowski to introduce the concept of unified space-time as a framework for physics, rather 
than separate newtonian space and time. 

It may strike the reader that there exist close similarities between this description of fields mediating 
electro-magnetic interactions and the decription of space-time geometry mediating gravitational 
interactions. There is indeed a field-theoretic approach to General Relativity that allows the reconstruction 
of GR starting from the theory of a symmetric tensor field \ct{veltman1975}. However, having the 
full theory of General Relativity at our disposal, it is easier and more convenient to reverse the 
procedure and derive the linear theory of a symmetric tensor field as a limit of GR for small-curvature 
perturbations in a Minkowski space-time. According to this procedure gravity can be described in the 
limit of weak fields in a flat space-time by a Lorentz-covariant tensor field $h_{\mu\nu} = h_{\nu\mu}$ 
subject to the following inhomogenous linear field equation\footnote{Unless stated otherwise we use 
natural units in which $c = 1$.}:
\be
- \Box\, h_{\mu\nu} + \der_{\mu} \der^{\lb} h_{\nu\lb} + \der_{\nu} \der^{\lb} h_{\mu\lb} 
 - \der_{\mu} \der_{\nu} h^{\lb}_{\;\,\lb} + \eta_{\mu\nu} \lh \Box\, h^{\lb}_{\;\,\lb} 
 - \der^{\kg} \der^{\lb} h_{\kg\lb} \rh = \kg T_{\mu\nu}.
\label{2.1}
\ee
Here $T_{\mu\nu}$ is a symmetric tensor representing the source of the field $h_{\mu\nu}$ and $\kg$ 
is a coupling constant defining the strength of the interaction between the field $h_{\mu\nu}$ and the 
source $T_{\mu\nu}$. For consistency the divergence of the tensor $T_{\mu\nu}$ has to vanish:
\be
\der^{\mu} T_{\mu\nu} = 0.
\label{2.2}
\ee
This is an important point to be addressed in detail in the following. First we turn to General Relativity and 
the Einstein equations, and show how to obtain and interpret eq.\ (\ref{2.1}) as its Minkowski limit.

\section{General Relativity \label{sec:3}} 

According to General Relativity gravity is an effect of space-time geometry. The mathematical ingredients 
of this geometry are the metric, the connection and the curvature. We briefly introduce these concepts 
here with the additonal aim to introduce our notation; more details can be found in the literature 
\ct{weinberg1972, hartle2003}. 

\begin{figure}[t]
\sidecaption[t]
\includegraphics[scale=.3]{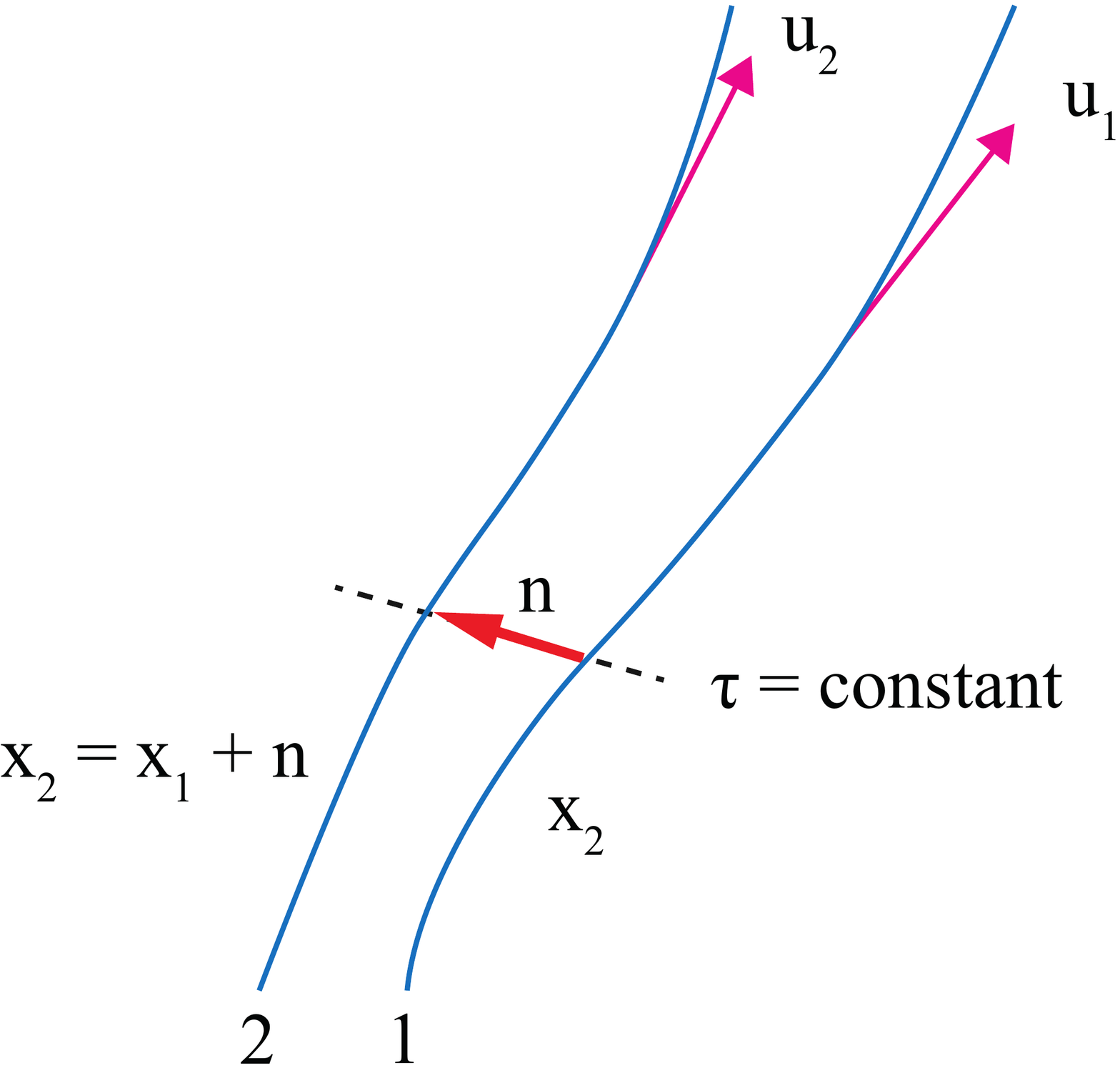}
%
%
\caption{Geodesic deviation as a measure of curvature.}
\label{fig:3.1}       
\end{figure}

The metric $g$ is the tool which allows us to convert co-ordinate measurements into physical space and 
time intervals. In particular a proper-time interval measured in a local rest frame can be expressed in 
terms of a quadratic form of co-ordinate intervals:
\be
- d\tau^2 = g_{\mu\nu}(x) dx^{\mu} dx^{\nu}.
\label{3.1}
\ee
Proper time is used to parametrize time-like curves $x^{\mu}(\tau)$, in particular the world-lines 
of massive particles. The unit tangent vector of the world-line is the 4-velocity:
\be
u^{\mu} = \dot{x}^{\mu}, \hs{2} g_{\mu\nu} u^{\mu} u^{\nu} = - 1.
\label{3.2}
\ee
As usual the overdot denotes an ordinary proper-time derivative. The inverse of the metric is used 
very often as well, and is distinguished by using superscripts for its components:
\be
g^{\mu\lb} g_{\lb\nu} = \del_{\nu}^{\mu}.
\label{3.3}
\ee
The connection is a quantity defining geodesics, the world-lines of particles in free fall. On such 
world-lines the 4-velocity only changes by parallel transport:
\be
D_{\tau} u^{\mu} = \dot{u}^{\mu} + \Gam_{\lb\nu}^{\;\;\;\mu} u^{\lb} u^{\nu} = 0.  
\label{3.4}
\ee
Here $\Gam$ is the Riemann-Christoffel connection and $D_{\tau}$ is called the covariant derivative 
along the world-line. The covariant derivative and the connection are defined in such a way that the 
metic itself is covariantly constant:
\be
\nb_{\lb} g_{\mu\nu}(x) = \der_{\lb} g_{\mu\nu} - \Gam_{\lb\mu}^{\;\;\;\kg} g_{\kg\nu} 
 - \Gam_{\lb\nu}^{\;\;\;\kg} g_{\mu\kg} = 0,
\label{3.5}
\ee
which implies that we can identify the connection with a linear combination of gradients of the metric:
\be
\Gam_{\mu\nu}^{\;\;\;\lb} = \frac{1}{2}\, g^{\lb\kg} \lh \der_{\mu} g_{\nu\kg} + \der_{\nu} g_{\mu\kg} 
 - \der_{\kg} g_{\mu\nu} \rh.
\label{3.6}
\ee
Because of this property it is clear that in free fall (on time-like geodesics) the metric changes only 
by parallel transport as well:
\be
\ld D_{\tau} g_{\mu\nu} \right|_{x(\tau)} = \dot g_{\mu\nu} - u^{\lb} \Gam_{\lb\mu}^{\;\;\;\kg} g_{\kg\nu}
 - u^{\lb} \Gam_{\lb\nu}^{\;\;\;\kg} g_{\mu\kg} = u^{\lb} \nb_{\lb} g_{\mu\nu} = 0. 
\label{3.7}
\ee
Finally the curvature measures the relative acceleration of geodesics; it is encoded in the Riemann 
tensor 
\be
R_{\mu\nu\kg}^{\;\;\;\;\;\;\;\lb} = \der_{\mu} \Gam_{\nu\kg}^{\;\;\;\lb} - \der_{\nu} \Gam_{\mu\kg}^{\;\;\;\lb}
 - \Gam_{\mu\kg}^{\;\;\;\sg} \Gam_{\nu\sg}^{\;\;\;\lb} + \Gam_{\nu\kg}^{\;\;\;\sg} \Gam_{\mu\sg}^{\;\;\;\lb}.
\label{3.8}
\ee
Suppose we have two particles on world lines $x_1(\tau)$ and $x_2(\tau) = x_1(\tau) + n(\tau)$;
see fig.\ \ref{fig:3.1}. For small separation $n$ can be replaced by the tangent vector of the geodesic
cutting the two world-lines in points of equal proper time $\tau$. The second derivative of this vector 
w.r.t.\ proper time then represents the proper acceleration between the particles on the two world-lines. 

Again defining its covariant derivative by parallel transport:
\[
D_{\tau} n^{\mu} = \dot{n}^{\mu} + u^{\lb} \Gam_{\lb\nu}^{\;\;\;\mu} n^{\nu}, 
\]                                  
by some straightforward algebra it can be established that up to quadratic corrections $n$ satisfies 
the condition 
\be
\hs{1}
 D_{\tau}^2\, n^{\mu} = R_{\kg\nu\lb}^{\;\;\;\;\;\;\;\mu}\, u^{\kg} u^{\lb} n^{\nu}.
\label{3.9}
\ee
Thus the components of the Riemann tensor at some point in space-time can be measured by 
determining the proper acceleration between various geodesics passing through that point. 
Obviously if the curvature vanishes there will be no relative acceleration, and the world-lines 
are straight as expected in flat space.

From the Riemann tensor one can construct the Ricci tensor and the Riemann scalar by contraction: 
\be
R_{\mu\nu} = R_{\nu\mu} = R_{\mu\lb\nu}^{\;\;\;\;\;\;\;\,\lb}, \hs{2} R = R_{\lb}^{\;\;\lb}.
\label{3.10}
\ee
Like the metric, the Ricci tensor is symmetric. Therefore the Einstein equations can fix the local 
geometry of space-time by relating the Ricci tensor directly to the energy-momentum distributions 
of matter and radiation as described by the symmetric energy-momentum tensor $T_{\mu\nu}$:
\be
G_{\mu\nu} \equiv R_{\mu\nu} - \frac{1}{2}\, g_{\mu\nu} R = - \kg^2 T_{\mu\nu},
\label{3.11}
\ee
where the constant of proportionality is related to Newtons constant of gravity: 
\be
\kg^2 = \frac{8\pi G}{c^4} \simeq 2.1 \times 10^{-41}\, \mbox{kg$^{-1}$ m$^{-1}$ s$^2$}.
\label{3.11.1}
\ee
Equation (\ref{3.11}) provides 10 independent inhomogeneous partial differential equations 
for the 10 independent components of the metric. 

The simplest solution of the Einstein equations in empty space is Minkowski space-time. 
It has a constant metric which can be taken in the form 
\be
g_{\mu\nu} = \eta_{\mu\nu} = \lh \ba{cccc} -1 & \;0 & \;0& \;0 \\
                                                                    0 & \;1 & \;0 & \;0 \\
                                                                    0 & \;0 & \;1 & \;0 \\
                                                                    0 & \;0 & \;0 & \;1 \ea \rh, 
\label{3.12}
\ee
corresponding to pseudo-euclidean co-ordinates in which the proper-time element reads
\be
- d\tau^2 = \eta_{\mu\nu} dx^{\mu} dx^{\nu} = - dt^2 + dx^2 + dy^2 + dz^2.          
\label{3.13}
\ee
As mentioned above this space-time has no curvature: the Riemann tensor vanishes, and 
geodesics are straight lines. 

Next we consider geometries with non-vanishing but small curvature. The metric of such a 
space-time will be close to the Minkowski metric (\ref{3.11}) and we expand it as 
\be
g_{\mu\nu} = \eta_{\mu\nu} + 2 \kg h_{\mu\nu}.
\label{3.14}
\ee
As the metric itself is dimensionless, the constant $2\kg$ is introduced to give $h_{\mu\nu}$ 
the canonical dimensions of a tensor field in 4-dimensional space-time. We can now compute 
the connection and the Ricci tensor under the assumption $\| 2 \kg h_{\mu\nu} \| \ll 1$. This 
guarantees that the geometry does not deviate strongly from Minkowski space-time. Therefore 
in calculating other geometrical quantities we may restrict ourselves to expressions linear 
in $h_{\mu\nu}$; in this limit the connection becomes 
\be
\Gam_{\mu\nu}^{\;\;\;\lb} \simeq \kg \lh \der_{\mu} h^{\;\;\,\lb}_{\nu} + \der_{\nu} h^{\;\;\,\lb}_{\mu}
 - \der^{\lb} h_{\mu\nu} \rh + {\cal O}(\kg^2),
\label{3.15}
\ee
where the switch from upper- to lower-index vectors and tensors has been made using the 
Minkowski metric. Similarly the Ricci tensor and the Riemann scalar take the form 
\be
\ba{l} 
\dsp{ R_{\mu\nu} \simeq \kg \lh \Box h_{\mu\nu} - \der_{\mu} \der_{\lb} h_{\nu}^{\;\;\,\lb}
 - \der_{\nu} \der_{\lb} h_{\mu}^{\;\;\,\lb} + \der_{\mu} \der_{\nu} h_{\lb}^{\;\;\,\lb} \rh
 + {\cal O}(\kg^2), }\\
  \\
\dsp{ R = 2 \kg \lh \Box h^{\lb}_{\;\,\lb} - \der^{\kg} \der^{\lb} h_{\kg\lb} \rh. }
\ea
\label{3.16}
\ee
Here the box operator represents the usual d'alembertian in Minkoswki space-time:
\be
\Box = \der_{\lb} \der^{\lb} =  - \der_t^2 + \der_x^2 + \der_y^2 + \der_z^2.
\label{3.17}
\ee
From this result it follows that the Einstein equations expanded to first order in $\kg$ 
reduce to equation (\ref{2.1}). Thus we see that this linear field equation for the symmetric 
tensor field $h_{\mu\nu}$ describes the fluctuations of space-time geometry close to flat 
Minkowski geometry. 
                                                 
\section{Gravitational Waves \label{sec:4}}

An important property of eq.\ (\ref{2.1}) for the symmetric tensor field is its invariance 
under local gauge transformations
\be
h_{\mu\nu} \rightarrow h'_{\mu\nu} = h_{\mu\nu} + \der_{\mu} \xi_{\nu} + \der_{\nu} \xi_{\mu}.
\label{4.1}
\ee
From the geometrical point of view of General Relativity these gauge transformations represent 
the reduction of general co-ordinate transformations to the theory of small fluctutations of 
Minkowski geometry. From the  point of view of tensor field theory in Minkowski space-time
the gauge invariance guarantees that the 10 components of the field actually describe only
two propagating physical modes and that as a consequence a number of components of 
$h_{\mu\nu}$ are redundant. Let us see in more detail how this works. 

First we are going to simplify the field equation by considering its trace; this results in 
\be
\Box\, h^{\lb}_{\;\;\lb} - \der^{\kg} \der^{\lb} h_{\kg\lb} = \frac{\kg}{2}\, T^{\lb}_{\;\;\lb}.
\label{4.2}
\ee
Therefore we can rewrite the tensor field equation in the form
\be
\lh \Box h_{\mu\nu} - \der_{\mu} \der_{\lb} h_{\nu}^{\;\;\,\lb} - \der_{\nu} \der_{\lb} h_{\mu}^{\;\;\,\lb} +
 \der_{\mu} \der_{\nu} h_{\lb}^{\;\;\,\lb} \rh = - \kg \lh T_{\mu\nu} - \frac{1}{2}\, \eta_{\mu\nu} T^{\lb}_{\;\;\lb} \rh.
\label{4.3}
\ee
As we have only rewritten the equation by grouping the elements differently the equation is still 
invariant under the gauge transformations (\ref{4.1}). We now use this freedom to impose additional
conditions on the tensor field. For example, we can perform a gauge transformation with parameters 
$\xi_{\mu}$ which are solutions of the equation 
\be
\Box\, \xi_{\mu} = - \der^{\lb} h_{\lb\mu} + \frac{1}{2}\, \der_{\mu} h^{\lb}_{\;\;\lb}. 
\label{4.4}
\ee
Then the transformed field satisfies both the field equation (\ref{4.3}) and in addition the gauge 
condition
\be
\der^{\lb} h'_{\lb\mu} - \frac{1}{2}\, \der_{\mu} h^{\prime\,\lb}_{\;\;\;\lb} = 0.
\label{4.5}
\ee
Moreover the field equation (\ref{4.3}) then simplifies further to 
\be
\Box\, h'_{\mu\nu} = - \kg \lh T_{\mu\nu} - \frac{1}{2}\, \eta_{\mu\nu} T^{\lb}_{\;\;\lb} \rh.
\label{4.6}
\ee
In particular in empty space the equation reduces to the massless wave equation
\be
\Box\, h'_{\mu\nu} = 0,
\label{4.7}
\ee
showing that the free tensor field describes gravitational waves propagating with the 
speed of light in Minkowski space-time.

Next we recall a mathematical theorem stating that Poincar\'{e} invariance (invariance under translations 
and Lorentz transformations) in 4-dimensional space-time implies that massless waves of vector and 
tensor fields possess only 2 transverse polarisation modes. This is actually well-known in the case 
of electro-magnetic waves, as the electric and magnetic fields are always directed at right angles to the 
direction of propagation. Here we derive the corresponding result for the symmetric tensor field. In brief 
the argument uses the observation that eq.\ (\ref{4.4}) specifies the gauge parameters only up to solutions 
of the homogeneous equation 
\be
\Box\, \xi_{\mu} = 0.
\label{4.8}
\ee
Gauge transformations of this type can be used to eliminate still more components of the field. For 
example, one can eliminate the trace with the result that also the 4-dimensional divergence of the 
field vanishes:
\be
h^{\prime \lb}_{\;\;\;\lb} = 0 \hs{.5} \Rightarrow \hs{.5} \der^{\lb} h'_{\lb\mu} = 0. 
\label{4.9}
\ee
The full analysis is most conveniently done in momentum space. Thus we expand the field in 
plane waves 
\be
h'_{\mu\nu}(x) = \int \frac{d^4 k}{(2\pi)^2}\, \ve_{\mu\nu}(k)\, e^{-i k \cdot x}, 
\label{4.10}
\ee
where the reality of $h'_{\mu\nu}(x)$ requires $\ve_{\mu\nu}^*(k) = \ve_{\mu\nu}(-k)$.
Then eqs.\ (\ref{4.5}) and (\ref{4.7}) are replaced by 
\be
k^{\lb} \ve_{\lb\mu} = \frac{1}{2}\, k_{\mu} \ve^{\lb}_{\;\;\lb}, \hs{2} k^2 \ve_{\mu\nu}(k) = 0, 
\label{4.11}
\ee
whilst the residual gauge transformations (\ref{4.1}), (\ref{4.8}) imply freedom to make redefinitions 
\be
\ve_{\mu\nu}(k) \rightarrow \ve'_{\mu\nu}(k) + k_{\mu} a_{\nu}(k) + k_{\nu} a_{\mu}(k), 
\label{4.12}
\ee
provided the momenta $k_{\mu}$ take values on the light-cone $k^2 = 0$. Now clearly also 
eq.\ (\ref{4.11}) allows non-zero solutions for $\ve_{\mu\nu}$ only if the momenta are on the light-cone. 
Therefore we can write
\be 
\ve_{\mu\nu}(k) = e_{\mu\nu}(\bfk) \del(k^2), \hs{2} e_{\mu\nu}^*(\bfk) = e_{\mu\nu}(-\bfk),
\label{4.13}
\ee
such that 
\be
k^{\mu} = (\pm \og, \bfk), \hs{2} \og = \sqrt{\bfk^2}.
\label{4.14}
\ee
Substitution of this expression for $\ve_{\mu\nu}(k)$ in the plane-wave expansion (\ref{4.10}) 
allows us to perform the integral over the time component $k^0$, with the result 
\be
h_{\mu\nu}(x) = \int \frac{d^3 k}{8 \pi^2 \og} e_{\mu\nu}(\bfk) \lh e^{- i (\bfk \cdot \bfx - \og t)}
 + e^{- i (\bfk \cdot x + \og t)} \rh.
\label{4.15}
\ee
The two terms arise from the two solutions for $k^0$ in eq.\ (\ref{4.14}). In the second term we can now 
replace the integration variable $\bfk \rightarrow - \bfk$ and use the complex conjugation condition on 
$e_{\mu\nu}$ in eq.\ (\ref{4.13}) to finally get 
\be
h_{\mu\nu}(x) = \int \frac{d^3 k}{8 \pi^2 \og} \lh e_{\mu\nu}(\bfk) e^{- i (\bfk \cdot \bfx - \og t)} + 
 e^*_{\mu\nu}(\bfk) e^{i (\bfk \cdot \bfx - \og t)} \rh.
\label{4.16}
\ee
In addition the gauge condition (\ref{4.11}) for the amplitude $\ve_{\mu\nu}$ implies after 
splitting space- and time-components and the replacement $k^0 = - k_0 = \og$:
\be
k_j e_{j0} = \frac{\og}{2} \lh e_{00} + e_{jj} \rh, \hs{1} 
 \lh \del_{ij} - \frac{k_i k_j}{\og^2} \rh e_{j0} = \frac{1}{\og} \lh  k_i e_{jj} - k_j e_{ji} \rh.
\label{4.17}
\ee
Here and in the following we extend the summation convention for repeated indices to apply 
also for 3-dimensional space-like indices $i,j = (1,2,3)$. 

Recall that the residual gauge transformations (\ref{4.12}) are also defined on the light-cone
$k^2 = 0$. On the amplitudes $e_{\mu\nu}$ they act as 
\be
\ba{l}
e'_{00} = e_{00} - 2 \og a_0, \hs{1} e'_{i0} = e_{i0} + k_i a_0 - \og a_i, \hs{1} 
e'_{ij} = e_{ij} + k_i a_j + k_j a_i.
\ea
\label{4.18}
\ee
If we make the following choice for the gauge parameters:
\be
a_0 = \frac{1}{2\og}\, e_{00}, \hs{1} a_i = \frac{1}{\og}\, e_{i0} + \frac{k_i}{2\og^2}\, e_{00},
\label{4.19}
\ee
it is straightforward to show with the help of the first equation (\ref{4.17}) that 
\be
e'_{00} = e'_{i0} = e'_{jj} = 0. 
\label{4.20}
\ee
By the second equation (\ref{4.17}) it follows in addition that 
\be
k_j e_{ji} = 0. 
\label{4.21}
\ee
In summary, starting from the field equation we can by mere gauge transformations restrict the 
plane-wave solutions for the tensor field to the light-cone and make them purely spatial, traceless 
and transverse. For example if the 3-momentum points in the $z$-direction: $\bfk = (0, 0, k)$,  
the amplitude can be decomposed as 
\be
e_{\mu\nu}(\bfk) = A_+(k) \sg^+_{\mu\nu} + A_{\times}(k) \sg^{\times}_{\mu\nu}, 
\label{4.22}
\ee
where $A_{+,\times}(k)$ represent the individual mode-amplitudes contributing to the Fourier integral 
and the polarisation tensors $\sg^{+,\times}$ are of the form 
\be
\sg^+ = \lh \ba{cccc} 0 & \;0 & \;0 & \;0 \\
                                 0 & \; 1 & \;0 & \;0 \\
                                 0 & \;0 & -1 & \;0 \\
                                 0 & \;0 & \;0 & \;0 \ea \rh, \hs{1}
\sg^{\times} = \lh \ba{cccc} 0 & \;0 & \;0 & \;0 \\
                                 0 & \;0 & \;1 & \;0 \\
                                 0 & \;1 & \;0 & \;0 \\
                                 0 & \;0 & \;0 & \;0 \ea \rh.                                 
\label{4.23}
\ee

\begin{figure}[b]
\includegraphics[scale=.22]{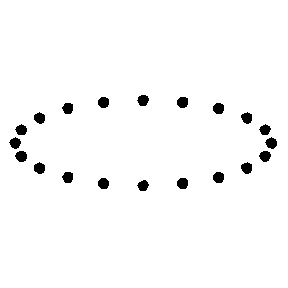} \hs{1} \includegraphics[scale=.22]{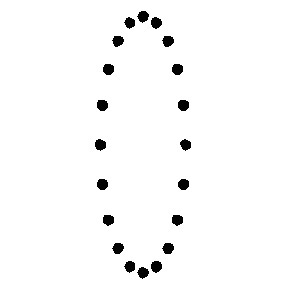}
\hs{2} \includegraphics[scale=.28]{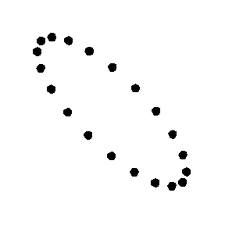}  \hs{1} \includegraphics[scale=.28]{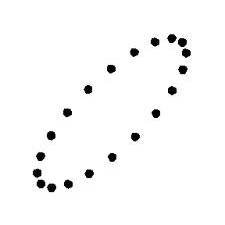}
%
%
\caption{Effect of the $+$- (left) and $\times$- (right) polarization modes of a gravitational 
wave on a ring of test masses in the transverse plane.}
\label{fig:4.1}       
\end{figure}
These polarization modes can be easily visualized. The $+$-mode defines a change in the
metric components $g_{xx}$ and $g_{yy}$ of opposite sign. Hence when physical distances 
in the $x$-direction increase, those in the $y$-direction decrease by the same amount, and
vice versa. The $\times$-mode changes the distances at 45$^\circ$ to the $x$- and $y$-axes 
in a similar way. If one imagines a ring of test masses in a plane perpendicular to the direction
of the wave, this ring will deformed by the two polarization modes of a wave as sketched in 
figure \ref{fig:4.1}.

\section{Energy considerations \label{sec:5}}

With the gauge choice (\ref{4.5}) the original wave equation (\ref{2.1}) reduces to 
\be
\Box\, h_{\mu\nu} - \frac{1}{2}\, \eta_{\mu\nu} \Box\, h^{\lb}_{\;\;\lb} = - \kg T_{\mu\nu}.
\label{5.1}
\ee
This equation can actually be derived from a hamiltonian as we now show. This hamiltonian 
is an expression in the fields $h_{\mu\nu}$ and conjugate momenta $\pi_{\mu\nu}$:
\be
H = \int d^3 x \left[ \frac{1}{2}\, \pi_{\mu\nu}^2 - \frac{1}{4}\, \pi^{\lb\,2}_{\;\;\lb} + \frac{1}{2} \lh \nb h_{\mu\nu} \rh^2 
 - \frac{1}{4} \lh \nb h^{\lb}_{\;\;\lb} \rh^2 - \kg h^{\mu\nu} T_{\mu\nu} \right].
\label{5.2}
\ee
The standard Hamilton equations of motion read
\be
\dot{h}_{\mu\nu} = \frac{\del H}{\del \pi^{\mu\nu}}, \hs{2}
\dot{\pi}_{\mu\nu} = - \frac{\del H}{\del h^{\mu\nu}}.
\label{5.3}
\ee
Using the expression (\ref{5.2}) this gives:
\be
\dot{h}_{\mu\nu} = \pi_{\mu\nu} - \frac{1}{2}\, \eta_{\mu\nu} \pi^{\lb}_{\;\;\lb}, \hs{1}
\dot{\pi}_{\mu\nu} = \Del h_{\mu\nu} - \frac{1}{2}\, \eta_{\mu\nu} \Del h^{\lb}_{\;\;\lb} + \kg T_{\mu\nu}.
\label{5.4}
\ee
Combining these equations one gets back eq.\ (\ref{5.1}). Now consider some volume $V$ in space;
the integral (\ref{5.2}) restricted to this volume at a fixed time $t$ may be considered as the total energy 
$E_V$ of the field in that volume. Its variation in time is 
\be
\ba{lll}
\dsp{ \frac{d E_V}{dt} }& = & \dsp{ 
  \frac{d}{dt} \int_V d^3x \left[ \frac{1}{2}\, \pi_{\mu\nu}^2 - \frac{1}{4}\, \pi^{\lb\,2}_{\;\;\lb} 
 + \frac{1}{2} \lh \nb h_{\mu\nu} \rh^2 - \frac{1}{4} \lh \nb h^{\lb}_{\;\;\lb} \rh^2 - \kg h^{\mu\nu} T_{\mu\nu} \right] }\\
 & & \\
 & = & \dsp{ \int_V d^3 x \left[ \nb \cdot \lh \pi^{\mu\nu} \nb h_{\mu\nu} \rh - \kg h^{\mu\nu} \dot{T}_{\mu\nu} \right] }\\
 & & \\
 & = & \dsp{ \oint_{\der V} d^2 \sg\, \pi^{\mu\nu} \nb_n h_{\mu\nu} - \kg \int_V d^3 x\, h^{\mu\nu} \dot{T}_{\mu\nu}. }  
\ea 
\label{5.5}
\ee
Here $\der V$ is the boundary surface of the volume $V$, $d^2 \sg$ is a surface element of the boundary 
and $\nb_n$ is the normal component of the gradient to that surface element. The physical interpretation of 
this equation is, that the energy inside the volume changes only by a flux of gravitational radiation through 
the boundary, up to changes in the material energy-momentum density inside. In particular for stationary 
sources $\dot{T}_{\mu\nu} = 0$ the energy can only change by radiation flowing through the surface $\der V$.

The energy flux itself is given by 
\be 
\Fg_i =  \pi^{\mu\nu} \nb_i\, h_{\mu\nu} = \dot{h}^{\mu\nu} \nb_i\, h_{\mu\nu} 
 - \frac{1}{2}\, \dot{h}^{\nu}_{\;\;\nu} \nb_i\, h^{\mu}_{\;\;\mu}. 
\label{5.6}
\ee
This establishes a quantitative relation between energy flux and amplitude of gravitational waves. For 
example, if we take a plane wave in the $z$-direction like in eq.\ (\ref{4.22}), (\ref{4.23}) they generate a 
dimensionless metric fluctuation 
\be
\del g_{ij} = 2 \kg h_{ij} = \mbox{Re}\, a_{ij} e^{2\pi i f (t - z/c)},
\label{5.7}
\ee 
and the energy flux in the $z$-direction per unit of area $A$ is
\be
\Fg_z = \frac{dE}{dA dt} = \frac{\pi c^3 f^2}{4G} \lh a_+^2 + a_{\times}^2 \rh,
\label{5.8}
\ee
where we have restored the the factors of $c$ so we can express the quantities in SI units. For example
an energy flux of $\Fg_z = 1$ mW/m$^2$ for a single polarisation mode at a frequency $f = 100$ Hz 
corresponds to an amplitude
\be
a_{+,\times} \simeq 10^{-21}.
\label{5.9}
\ee
From this example one can learn two things: \\
({\em i}) The amplitude of metric deformations by gravitational waves are extremely small; the metric 
deformation for the wave in our example implies that the distance between the mirrors of an interferometer 
with an arm length of 1 km changes by 1/100th of the diameter of a proton. \\
({\em ii}) Space is extremely stiff: it takes very large energy densities to create even small deviations from 
flatness. For comparison: in spite of the very small amplitude the energy flux in our example corresponds 
to a spectral brightness of $10^{21}$ Jy,  which is 15 - 25 orders of magnitude larger than those customary 
in radio astronomy. Therefore only strong astrophysical sources may have detectable effects.

\section{Gravitational Wave Emission \label{sec:6}}

Suppose that we have a source consisting of one or several masses like a pulsar, a supernova or a 
binary star system with energy-momentum tensor $T_{\mu\nu}$. We would like to compute the 
gravitational-wave flux observed at a large distance $r$ far away form the source itself. We imagine 
a large spherical surface with radius $r$ centered on the source but sufficiently far away that 
$T_{\mu\nu}$ vanishes on the surface. We can then use eq.\ (\ref{5.5}) with spherical surface element 
\be
d^2 \sg = r^2 d \Og \equiv r^2 \sin \thg\, d\thg d\vf,
\label{6.1}
\ee
to compute the energy flux through the spherical angle $d\Og$:
\be
\frac{dE}{d\Og dt} = 
 r^2  \lh \dot{h}^{\mu\nu} \nb_r h_{\mu\nu} - \frac{1}{2}\, \dot{h}^{\nu}_{\;\;\nu} \nb_r h^{\mu}_{\;\;\mu} \rh. 
\label{6.2}
\ee
Clearly contributions to the amplitude falling off faster than $1/r$ will not contribute at large $r$, hence 
they will be neglected in this computation. Now in view of the field equation (\ref{5.1}) it is actually 
convenient to redefine the field variables:
\be
\uh_{\,\mu\nu} \equiv h_{\mu\nu} - \frac{1}{2}\, \eta_{\mu\nu} h^{\lb}_{\;\;\lb} \hs{1} 
\Leftrightarrow \hs{1} h_{\mu\nu} = \uh_{\,\mu\nu} - \frac{1}{2}\, \eta_{\mu\nu} \uh^{\lb}_{\;\;\lb}.
\label{6.3}
\ee
This new field variable now satisfies
\be
\Box\, \uh_{\,\mu\nu} = - \kg T_{\mu\nu}, \hs{1} \der^{\nu} \uh_{\,\nu\mu} = 0, \hs{1} 
\der^{\nu} T_{\nu\mu} = 0.
\label{6.4}
\ee
Conveniently the equation for the energy flux is the same in terms of $h_{\mu\nu}$ and
$\uh_{\mu\nu}$: 
\be
\frac{dE}{d\Og dt} = r^2 \lh \underline{\dot{h}}^{\mu\nu} \nb_r \uh_{\,\mu\nu} 
   - \frac{1}{2}\, \underline{\dot{h}}^{\nu}_{\;\;\nu} \nb_r \uh^{\mu}_{\;\;\mu} \rh. 
\label{6.5}
\ee
Equation (\ref{6.4}) has the standard retarded-time solution 
\be
\ba{lll}
\uh_{\mu\nu}(\bfx,t) & = & \dsp{ 
  - \frac{\kg}{4\pi} \int d^3x'\, \frac{T_{\mu\nu}(\bfx', t - |\bfx' - \bfx|)}{|\bfx' - \bfx|} }\\
 & & \\
 & \simeq & \dsp{ - \frac{\kg}{4\pi r} \int d^3x'\, T_{\mu\nu}(\bfx', t - r), }
\ea
\label{6.6}
\ee
modulo terms vanishing faster than $1/r$. Next we can argue that the time-components of 
the field don't contribute to the flux. Indeed as the divergence of $T_{\mu\nu}$ vanishes, 
the time derivative of $\uh_{0\mu}$ is turned into a boundary term which vanishes as well:
\be
\ba{lll}
\underline{\dot{h}}_{0\mu} & = & \dsp{ - \frac{\kg}{4\pi r} \int d^3x'\, \dot{T}_{0\mu} =
 -  \frac{\kg}{4\pi r} \int d^3x'\, \nb_j T_{j\mu} }\\
 & & \\
 & = & \dsp{ \oint d^2 \sg\, T_{n\mu} = 0. }
\ea
\label{6.7}
\ee
Thus we are left only with the spatial components $\uh_{ij}$.  Far away from the source they behave 
like free fields and therefore the contributions to the flux come from the components which are 
traceless and transverse. On the spherical surface with outward normal unit vector $\hat{\bfr}$ this implies 
\be
\uh_{ij} = - \frac{\kg}{4\pi r} \lh \del_{ik} - \hr_i \hr_k \rh  \lh \del_{jl} - \hr_j \hr_l \rh 
 \lh I_{kl} + \frac{1}{2}\, \del_{kl} \hr \cdot I \cdot \hr \rh,
\label{6.8}
\ee
where $I_{ij}$ is the traceless part of the retarded volume integral of $T_{ij}$: 
\be
\ba{l} 
\dsp{ I_{ij} = \int d^3x' \lh T_{ij} - \frac{1}{3}\, \del_{ij} T_{kk} \rh (\bfx', t-r). }
\ea
\label{6.9}
\ee
Indeed, this expression satisfies the conditions
\be
\uh_{jj} = 0, \hs{1} \hr_j \uh_{ji} = 0.
\label{6.9.1}
\ee
Now one can replace $T_{ij}$ in the integrand (\ref{6.9}) by the second moment of $T_{00}$ as follows:
\be
\int d^3 x'\, T_{ij}(\bfx',t-r) = \frac{1}{2}\, \dd{^2}{t^2} \int d^3x'\, x'_i x'_j\, T_{00}(\bfx',t-r).
\label{6.10}
\ee
The proof requires applying the vanishing divergence of the energy-momentum tensor twice:
\be
\dd{^2}{t^2} T_{00} = \dd{}{t} \nb_i T_{i0} = \nb_i \nb_j T_{ij},
\label{6.11}
\ee
and then performing two partial integrations. As a result we get
\be
I_{ij} = \frac{1}{2}\, \dd{^2}{t^2} \int d^3x' \lh x'_i x'_j - \frac{1}{3}\, \del_{ij}\, \bfx^{\prime\,2} \rh T_{00}(\bfx', t-r). 
\label{6.12}
\ee
Observe that for non-relativistic sources the energy density $T_{00}$ is dominated by the 
mass density, and $I_{ij}$ reduces to a second time-derivative of the quadrupole moment of 
the mass distribution:
\be
I_{ij} = \frac{1}{2}\, \dd{^2 Q_{ij}}{t^2}, \hs{1}
 Q_{ij} \simeq \int d^2x' \lh  x'_i x'_j - \frac{1}{3}\, \del_{ij}\, \bfx^{\prime\,2} \rh \rg(\bfx', t-r).
\label{6.13}
\ee
Therefore 
\be
\uh_{ij} = - \frac{\kg}{8\pi r} \lh \del_{ik} - \hr_i \hr_k \rh  \lh \del_{jl} - \hr_j \hr_l \rh \dd{^2}{t^2} 
 \lh Q_{kl} + \frac{1}{2}\, \del_{kl} \hr \cdot Q \cdot \hr \rh.
\label{6.14}
\ee
Finally in computing the energy flux we can use the fact that $Q$ depends only on $t - r$, and 
therefore
\be
\nb_r Q_{ij} = - \dd{Q_{ij}}{t}.
\label{6.14.1}
\ee
Hence the distant energy flux is given by
\be
\frac{dE}{d\Og dt} = - \frac{G}{8\pi} \left[ \ddd{Q}_{ij}\,{^{\hs{-.4}2}} - 2 \hr \cdot {\ddd{Q}}\,{^{\,2}} \cdot \hr 
  + \frac{1}{2} (\hr \cdot \ddd{Q} \cdot \hr)\,{^2} \right],
\label{6.15}
\ee
where the triple overdots denote the third derivative w.r.t.\ time and the minus sign signifies 
that energy is leaving the spherical volume. If the radiation is isotropic we can average over 
all directions, using 
\be
\langle \hr_i \hr_j \rangle = \frac{1}{3}\, \del_{ij}, \hs{1} 
\langle \hr_i \hr_j \hr_k \hr_l \rangle = \frac{1}{15} \lh \del_{ij} \del_{kl} + \del_{ik} \del_{jl} 
 + \del_{il} \del_{jk} \rh.
\label{6.16}
\ee
Then integrating the spherical angle $d\Og$ over the full range $4\pi$ and resoring the 
factors of $c$ the total flux equals
\be
\frac{dE}{dt} = - \frac{G}{5c^5} \ddd{Q}_{ij}\,{^{\hs{-.4}2}}. 
\label{6.17}
\ee

\section{Newtonian Binaries \label{sec:7}}

\begin{figure}[t]
\sidecaption[t]
\includegraphics[scale=.5]{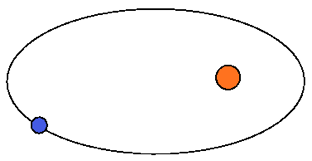}
%
%
\caption{Binary system of two well-separated stars dominated by newtonian gravity.}
\label{fig:7.1}       
\end{figure}

A straightforward and relevant application of the theory of gravitational-wave emission in 
Minkoswki space-time is the energy-loss of non-relativistic binary star systems, for example
consisting of white dwarf and neutron stars in the newtonian limit as sketched in figure~\ref{fig:7.1}. 
We treat the stars as point masses $m_{1,2}$ with positions $\bfr_{1,2}$. Writing 
$M = m_1 + m_2$ and $\bfr = \bfr_1 - \bfr_2$ Newtons law of gravity provides the 
equation of motion 
\be
\ddot{\bfr} = - \frac{GM}{r^3}\, \bfr. 
\label{7.1}
\ee
For simplicity we consider circular orbits $r =$ constant, with the center of mass at rest. 
Conservation of angular momentum guarantees that the joint orbit lies in a plane, which 
we take to be the equatorial plane. Then the motion of the individual masses can be 
parametrized by
\be
\bfr_1 = \frac{m_2 r}{M} \lh \cos \og t, \sin \og t, 0 \rh, \hs{1}
\bfr_2 = - \frac{m_1 r}{M} \lh \cos \og t, \sin \og t, 0 \rh,
\label{7.2}
\ee
where the angular frequency is given in accordance with Keplers law by
\be
\og^2 = \frac{GM}{r^3}.
\label{7.3}
\ee
For point masses the quadrupole moment is a sum of terms rather than an integral:
\be
\ba{lll}
Q_{ij} & = & \dsp{ 
  m_1 \lh r_{1\,i} r_{1\,j} - \frac{1}{3} \del_{ij}\, \bfr_1^2 \rh + m_2 \lh r_{2\,i} r_{2\,j} - \frac{1}{3} \del_{ij}\, \bfr_2^2 \rh }\\
 & & \\
 & = & \dsp{ \frac{\mu r^2}{2} \lh \ba{ccc} \cos 2 \og t + \frac{1}{3} & \sin 2 \og t & 0 \\
                                                        \sin 2 \og t & - \cos 2 \og t + \frac{1}{3} & 0 \\
                                                         0 & 0 & - \frac{2}{3} \ea \rh, }
\ea
\label{7.4}
\ee
where $\mu = m_1 m_2/M$ is the reduced mass of the system. 
This expression for the quadrupole moment can now be substituted in eq.\ (\ref{6.17})
to give the energy loss as 
\be
\frac{dE}{dt} = - \frac{G}{5c^5}\, 32 \mu^2 r^4 \og^6 = - \frac{32 G^4}{5c^5} \frac{m_1^2 m_2^2 M}{r^5}.
\label{7.5}
\ee
Observe, that one can also write this as 
\be
\frac{dE}{dt} = - \frac{c^5}{5G} \lh \frac{\rg_1}{r} \rh^2 \lh \frac{\rg_2}{r} \rh^2 \frac{\rg_1 + \rg_2}{r},
\label{7.6}
\ee
where $\rg_i$ denotes the corresponding Schwarzschild radius 
\be
\rg_i = \frac{2Gm_i}{c^2}.
\label{7.7}
\ee
The results obtained here can be applied for example to the case of binary neutron stars or white dwarfs; 
for equal mass systems 
\be
\left| \frac{dE}{dt} \right| = \frac{2c^5}{5G} \lh \frac{\rg}{r} \rh^5.
\label{7.9}
\ee
Now if this the gravitational energy flux is measured at a large distance $R$ on average the flux
will be
\be
\Fg = \frac{1}{4\pi R^2} \left| \frac{dE}{dt} \right| = \frac{c^5}{10 \pi G R^2} \lh \frac{\rg}{r} \rh^5.
\label{7.10}
\ee
We can compare this with the flux formula (\ref{5.8}) for plane waves to get an estimate of the 
amplitude
\be
\Fg = \frac{\pi c^3 f^2}{4G} \left| a \right|^2 = \frac{c^5}{16 \pi G r^2 } \frac{\rg}{r} \left| a \right|^2,
\label{7.11}
\ee
where we have used the expression (\ref{7.3}) for the frequency:
\be
f^2 = \lh \frac{\og}{2\pi} \rh^2 = \frac{GM}{4\pi^2 r^3} = \frac{c^2}{4\pi^2 r^2} \frac{\rg}{r}.
\label{7.12}
\ee
Equating the results (\ref{7.10}) and (\ref{7.11}) we get for the amplitude 
\be
\left| a \right|^2 = \frac{8}{5} \lh \frac{r}{R} \rh^2 \lh \frac{\rg}{r} \rh^4.
\label{7.13}
\ee
For example two neutron stars of 1.4 solar mass each revolving around each other at a distance of
$1.5 \times 10^6$ km and observed from a distance of 6.4 kpc create gravitational waves of amplitude 
$|a| \sim 0.8 \times 10^{-22}$ and frequency $f \sim 0.5 \times 10^{-4}$ Hz.

\begin{figure}[t]
\includegraphics[scale=.4]{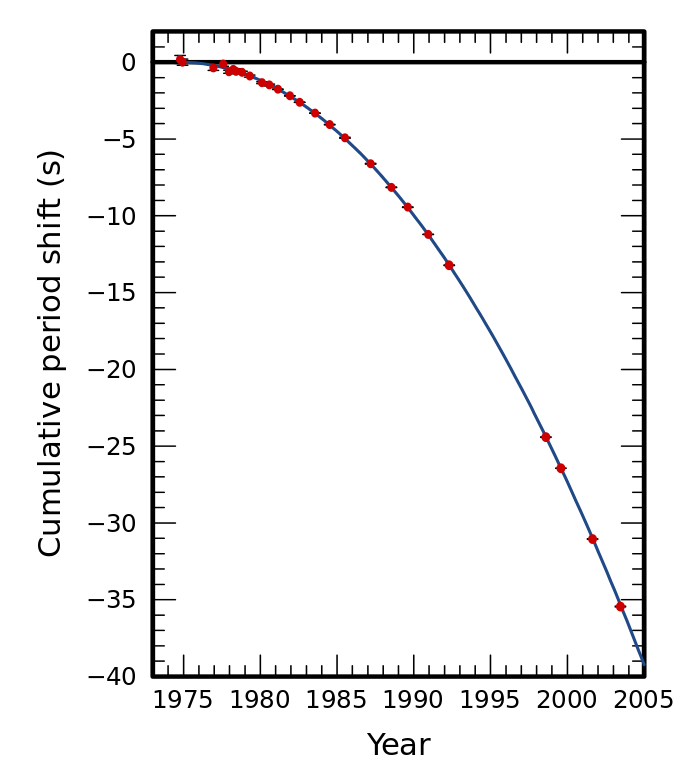}
%
%
\caption{Cumulative periastron shift of binary pulsar PSR 1913+16 \ct{weisberg-taylor2004}. 
The dots represent the measured shift, the curve the expectation from GR including emission
of gravitational waves.}
\label{fig:7.2}       
\end{figure}

These numbers are similar to those of the binary neutron star discovered in 1975 by Hulse 
and Taylor; they identified a pulsar which was part of a close binary system with a period of 
orbital revolution of about 7.5 hours. The masses of the stars were determined to be 
$m_1 = 1.39\, M_{\odot}$ and $m_2 = 1.44\, M_{\odot}$ and their distance to the solar system 
is $R = 6.4$ kpc.  Due to the emission of gravitational waves the orbit is expected to decay: 
the radius shrinks and the orbital period should decrease. As for this system the orbit is quite 
non-circular this change in period implies a shift in the periastron in addition to the one 
predicted by General Relativity for stable orbits around a star. Such a shift has been observed 
indeed over a period of more than 30 years, see figure \ref{fig:7.2}; it is in close quantitative 
agreement with the expectations for gravitational wave emission \ct{weisberg-taylor2004}. 
The challenge this observation and similar later ones set is of course to detect these gravitational 
waves directly. 

The result (\ref{7.6}) also provides theoretical limit on the energy output of two revolving compact 
masses in gravitational radiation: as $r$ can never become smaller than $\rg_1 + \rg_2$ we get
\be
\left| \frac{dE}{dt} \right| < \frac{c^5}{20 G} = 1.8 \times 10^{51}\, \mbox{W},
\label{7.8}
\ee
for two equal masses orbiting close to horizon distance. Apart from the fact that this limit 
presupposes the masses to be black holes, the estimate actually neglects all relativistic 
effects on the orbits as wel as radiation reaction. In practice the limit is actually more restrictive. 
Nevertheless the energy output can be enormous, up to 25 orders of magnitude larger than the 
present total energy output of the sun. 

\section{Observing gravitational waves \label{sec:8}} 

In view of the strong sources required and the small amplitudes involved the observation of gravitational
waves is a major challenge. Common strategies to detect gravitational waves use apparatus exchanging
energy with the wave field, either by absorption or emission. Two different strategies have been tried so far. 
The first one was the use of mechanical resonant oscillators, like solid bars or spheres. A second option is
to use laser interferometry, either on the surface of the earth or in space. In addition there are strategies to 
establish the effects of gravitational waves on astronomical or cosmological phenomena, like the motion of
pulsars or the imprint on the cosmic microwave background (CMB). We will briefly discuss some basic 
principles of these strategies. More detailed information is presented in the other lectures at this school
and in the literature \ct{weinberg1972,saulson1994,maggiore2008}. 
\vs{1}

\nit
{\em Resonant detectors} \\
The first attempts to detect gravitational waves were made by J.\ Weber 45 years ago \ct{weber1968}. 
He equipped two massive aluminum cylinders with accelerometers at the end surfaces. When a 
gravitational wave passes the apparatus, it will change its length by squeezing or stretching; if this 
is done in resonance with the fundamental vibration mode of the cylinder the energy of vibration can 
be changed. Subsequently a mechanical transducer can be used to amplify the change in amplitude 
to a detectable level. Modern versions of these mechanical detectors operate at cryogenic temperatures 
by cooling with liquid helium to reduce thermal noise. The best detectors have reached strain sensitivities 
of about $10^{-21}$ over a bandwidth of 1 Hz near the frequency of the fundamental mode, typically in 
the range 900-1000 Hz. In spite of Weber's original claims, no detection has ever been confirmed so far. 
\vs{1}

\nit
{\em Interferometers} \\
An interferometer consists of two --usually orthogonal-- arms and a beam of light which is split at the
point where the arms join to create two beams traveling along each arm. At the end of the arms there 
are mirrors reflecting the light beams back to the splitting point where the two beams can be compared
to find out differences in traveling time. The technical details are much more sophisticated, but this 
description suffices for an understanding of the principles. 

When a gravitational wave hits the interferometer the distances between the mirrors and the beam 
splitter change; the $+$-polarization mode with respect to the axes defined by the lay-out of the
interferometer, as sketched in figure \ref{fig:4.1}, will cause one arm to become shorter and the 
other one longer. As a result the light beams arrive back at the splitter after a round trip at 
different times. In addition the collision with the accelerated mirror at the time of reflection creates 
a change in wavelength. The upshot is a measurable difference in phase and amplitude of the 
recombined light beams from which information about the amplitude and frequency of the 
gravitational wave can be reconstructed. 

Large-scale interferometers presently operate in the USA: two LIGO detectors \ct{ligo}, and in 
Europe: VIRGO \ct{virgo} an GEO-600 \ct{geo600}. Similar devices are being planned elsewhere, 
e.g.\ in Japan and India. In the first round of operation LIGO and VIRGO reached typical strain 
sensitivities of $10^{-22}$ in the frequency domain between 50-1000 Hz. Presently upgrades of 
the instruments are underway to improve these sensitivities by yet another order of magnitude.

Although these broad-band frequency sensitivities are quite impressive, unfortunately 
interferometers are mostly insensitive to the associated $\times$-polar\-ization modes of  
gravitational-wave signals which create no difference in arrival times of the light beams. This 
also makes it very difficult to determine the polarization plane of the waves and 
their direction. Hence a single interferometer can detect a signal, but cannot identify its 
source. However by comparing the arrival times of signals at three interferometers not in
the transversal polarization plane of the wave, it does become possible to obtain information 
about the direction of the wave and thus of its source. Therefore the existing large-scale 
interferometers mentioned above all combine data to allow for better and more accurate 
identification of gravitational wave signals. 

Most resonant detectors actually suffer from the same problem of directional insensitivity, 
as the bar-type detectors also register a single polarization mode. An exception to this are 
spherical resonant detectors, like mini-Grail once operational at Leiden University \ct{mgrail}, 
which was equally sensitive to all polarization modes and could in principle determine the 
direction of propagation of the wave.
\vs{1}

\nit 
{\em Space based detectors} \\
Terrestrial interferometers also suffer from the problem that they cannot operate at frequencies
below about 20 Hz due to seismic noise which drowns any signal. In particular the gravitational 
waves emitted by the known binary pulsars, in the range of mHz and below, arre unobservable 
with existing instruments. That problem could be solved by building a free-floating interferometer 
in space. Plans for such an instrument exist at the European Space Agency. The eLISA project
\ct{elisa} aims to bring three satellites carrying laser equipment in an orbit following the earth 
around the sun. The distance between the satellites creating the arms of this interferometric type 
of instrument will be about 1 000 000 km and be sensitive to gravitational waves with amplitudes 
of $10^{-21}$ in the mHz range from more or less anywhere in the universe. A first test with 
instrumentation will be made during the Pathfinder mission to be launched in December 2015. 
The actual mission of eLISA itself is scheduled in the years after 2030. 
\vs{1}

\nit
{\em Pulsar timing} \\
A completely different approach to observe very low frequency gravitational waves is offered by the 
method of pulsar timing. Pulsars are rapidly spinning neutron stars emitting beams of radio-frequency
radiation rotating with the star. If the beam sweeps over the earth this results in extremely regular 
pulses of radiowaves. As some pulsars spin at rates of several tens to several hundreds cycles
per second, pulsars act as millisecond clocks, which are actually stable to one part in a million or 
better. Therefore pulsars can be used to keep time at the nanosecond level. Now the pulsars we
observe with such regularity are located in our galaxy, typically at distances of several hundreds of 
parsecs (1 pc = 3.1 light years). Therefore the travel time of the radiosignals is of the order of 
$10^{10}$ seconds. Thus in principle one can to determine variations in the travel time of the 
signal of 1 part in $10^{19}$ - $10^{20}$. By monitoring a sufficiently large set of pulsars 
(typically 20-50) with this accuracy to get statistically significant data on the relative motion of
the pulsars with respect to earth it may become possible to discover variations due to the passage 
of gravitational waves. This method has maximal sensitivity for waves in the frequency range of
$10^{-8}$ - $10^{-4}$ Hz, well below the eLISA range. It might be of importance to measure 
gravitational radiation from binary supermassive black holes, which can be created in the merging 
of galaxies. Several international research groups have started the timing of such pulsar networks; 
they collaborate in the International Pulsar Timing Array network \ct{ipta}. 
\vs{1}

\nit
{\em Polarization of the CMB} \\
The cosmic microwave background is a diffuse thermal background of photons originating 
from the time of formation of primordial neutral hydrogen and helium in the early universe,
when it was approximately 350 000 years old. Presently the average temperature of these
photons is 2.7 K which implies the spectrum peaks in the microwave range ($\sim$ 100 GHz).
The density fluctuations in primordial matter have been measured with great precision from the 
variations in temperature of the CMB over the sky. Gravitational waves from the early universe 
would create tensor fluctuations in the CMB, resulting in polarization patterns known as 
$B$-modes: patterns with non-vanishing curl. Typically gravitational waves of primordial 
origin visible $10^{10}$ years later in the CMB have frequencies of the order of $10^{-17}$ Hz,
again orders of magnitude below those searched for by the pulsar timing arrays. One 
experiment, BICEP2 located in Antarctica, has claimed to have detected evidence for $B$-mode 
polarization in the CMB \ct{bicep2-2014}, but this has not been confirmed by other observations 
like those made by the Planck satellite mission \ct{planck2013}. 

\section{Conclusions and outlook \label{sec:9}}

We can summarize our findings as follows. The linear abelian spin-2 field theory in Minkowski 
space-time is a first-order approximation to fluctuations of the space-time metric in a near-Minkowski 
space-time. The fluctuations have two transverse polarization states and propagate at the speed of 
light and are identified with gravitational waves. The waves can be sourced by a time-variable 
quadrupole moment of matter and the intensity of such waves has been calculated. Even for 
strong and massive astrophysical sources the resulting amplitudes of gravitational waves are 
extremely small and will be detectable only with great effort. 

We have discussed only the simplest case of gravitational waves in flat space-time. However, it
is possible to proceed in analogous ways to calculate wave propagation in curved space-times,
like cosmological geometries for an expanding universe or the strong-curvature regime of black 
holes \ct{maggiore2008,martel-poisson2005,koekoek-vholten2011}. Especially the merger phase 
of compact objects ---white dwarfs, neutron stars or black holes--- will produce strong and interesting 
wave signals from wich we can learn about gravity in the strong-field regime. Gravitational waves 
can even carry information about the earliest phase in the history of the universe before the universe 
became transparent to light or neutrinos. 

Indirect evidence for the emission of gravitational waves by binary neutron stars is quite strong by
now. It remains to observe this interesting phenomenon directly. Once the technology is in place
a new era in astronomy and astrophysics will open. 
\vs{3}

\begin{acknowledgement}
The work of the author is supported by the Foundation for Fundamental Reserach of Matter (FOM).
\end{acknowledgement}

\end{document}